\documentclass{aa}

\usepackage{natbib}

\newcommand{\fuse}{\emph{FUSE}}
\newcommand{\hst}{\emph{HST}}
\newcommand{\stis}{\emph{HST/STIS}}

\begin{document}

\title{Detection of a companion to the pulsating sdB
  Feige~48\thanks{Based on observations made with the NASA/ESA Hubble
  Space Telescope, which is operated by the Association of
  Universities for Research in Astronomy, Inc., under NASA contract
  NAS 5-26555. These observations are associated with program
  \#8635.}\fnmsep\thanks{Based on observations made with the NASA-CNES-CSA Far
  Ultraviolet Spectroscopic Explorer. FUSE is operated for NASA by the
  Johns Hopkins University under NASA contract NAS5-32985.}}

\author{S.~J. O'Toole
  \inst{1}
  \and U. Heber
  \inst{1}
  \and R.~A. Benjamin
  \inst{2}
}

\institute{Dr Remeis-Sternwarte, Astronomisches Institut der
  Universit\"at Erlangen-N\"urnberg, Sternwartstr. 7, Bamberg D-96049,
  Germany
  \and Department of Physics, University of Wisconsin-Whitewater, Whitewater,
  WI 53190 USA
}

\offprints{Simon O'Toole \email{otoole@sternwarte.uni-erlangen.de}}

\date{Received / Accepted}

\abstract{
We present the discovery of a binary companion to the pulsating sdB
Feige~48. Using HST/STIS and archival FUSE spectra, we measure a period of
0.376$\pm$0.003\,d and a velocity semi-amplitude of
28.0$\pm$0.2\,km\,s$^{-1}$. This implies that the companion star must either
be of very low mass, or the orbit is at high inclination. Combining
\emph{2MASS} fluxes, the lack of a reflection effect, results from
asteroseismology and a measurement of the rotation velocity of Feige~48, we
show that the orbital inclination must be $\le$11.4$^\circ$ and that the
unseen companion is a white dwarf with mass $\ge$0.46\,$M_\odot$. 
The implications of this discovery, and of
binarity amongst sdB pulsators, is then discussed in the context of recent
theoretical work on sdB formation. In particular we suggest that radial
velocity studies focus on sdB pulsators with no known companion, and that
asteroseismological studies of sdBs investigate a larger mass range than
previously considered in order to test formation models. 
}

\titlerunning{A companion to the pulsating sdB Feige~48}
\authorrunning{S.~J. O'Toole et al.}

\maketitle

\section{Introduction}
\label{sec:intro}

In recent years, interest in hot subdwarf B (sdB) stars has increased
dramatically. This is in large part due to the discovery of rapid oscillations
in some sdBs, but also because improving instrumentation has
allowed more detailed studies of their spectral properties. It is now
generally accepted that sdBs are associated with stars on the
extreme Horizontal Branch (EHB) \citep[e.g.,][]{Heber86, SBK94}. This
means they are low-mass stars ($\sim$0.5\,$M_\odot$) that burn helium
in their cores and have hydrogen envelopes too thin
($<$0.02\,$M_\odot$) to sustain nuclear burning. Stellar evolution
calculations by \citet{DRO93} predict that because of their thin
envelopes, sdBs will evolve directly to the white dwarf cooling track,
bypassing the Asymptotic Giant Branch.

While the next stages of sdB evolution appear to be known, the
question of sdB formation remains unanswered. There have been three
basic scenarios proposed: binary interaction involving Roche lobe
overflow \citep{MNG76}, single star evolution with strong mass loss
near the tip of the Red Giant Branch \citep{DCru96b}, and the merger
of two helium white dwarfs \citep{IT1986}. Significant progess has
been made with the recent work of \citet{HanI,HanII}, who studied
three possible binary formation channels: common envelope ejection,
stable Roche lobe overflow and the merger of two helium white
dwarfs. Using binary population synthesis, they created 12 simulations and
found one of them could satifactorily reproduce the observed characteristics
of sdBs ($T_{\mathrm{eff}}$, $\log~g$, binary fraction, orbital period
distribution, etc) when compared to radial velocity studies by
\citet{MHMN01} and \citet{M-RMM03}. \citet{RKL04} found a binary
fraction of 39\%, lower than predicted, while \citet{Lisker2004}, in
the largest homogenous study so far, compared their data to the simulations of
\citet{HanII} and found a different best-fit model.

The oscillations detected in some sdBs offer a way to test these
formation models. Importantly, pulsations are seen in sdBs with short-
and long-period binary companions, as well as in apparently single
sdBs. The properties of the V361\,Hya stars (the short-period,
$p$-mode sdB pulsators, originally called EC\,14026 stars) are
reviewed by \citet{Kilkenny2002}, while \citet{GFR03} presents the
detection of long-period, likely $g$-mode pulsations in somewhat
cooler sdBs (currently known as the PG\,1716+426 or ``Betsy'' stars).

Pulsations in the V361\,Hya star, Feige~48, were first detected by
\citet{ECpaperXI}. It is one of the coolest sdB pulsators in its
class, with $T_{\mathrm{eff}}$=29\,500\,K and $\log g$=5.50
\citep{HRW00}, and was discovered to be pulsating with five periods
between 344 and 379\,s. The oscillation amplitudes show dramatic
variation from night to night, which the authors suggested are due to
either changes in pulsation power, the beating of unresolved modes, or
a combination of the two. In a comprehensive follow-up study,
\citet{RKZ04} observed Feige~48 over 5 years, and found that the periods
measured by \citet{ECpaperXI} were resolved and that the amplitudes vary by at
least 30\%. A tentative model match to four out of five detected periods,
suggests that Feige~48 rotates with a period around 10\,h. The
$T_{\mathrm{eff}}$ and $\log g$ of this model were chosen to be consistent
with those derived from spectroscopy. Long-term monitoring by
\citeauthor{RKZ04} also provided an opportunity to measure the phase stability
of the pulsations. In particular, they were able to place constraints on the
period and mass of a companion to quite high precision, limiting any
stellar-mass companion to have a period shorter than 3 days. In another recent
study, \citet{CBF03} found a very good model match to all nine periods
detected by high S/N photometry using the CFHT. They derived a rotational
period of 9.58\,h, in good agreement with \citeauthor{RKZ04}

In this paper we present the detection of binary motion in Feige~48,
an sdB that was generally assumed to be single. We discuss the nature
of the companion, and implications of this detection for current
theories of sdB formation and pulsation.

\section{Observations}
\label{sec:obs}

Observations with \stis\ were made as part of an ongoing project to
measure metal abundances in pulsators and non-pulsators (proposal
\#8635, PI: U.\ Heber). Analysis of
these data will be presented in a future paper \citep[see][ for
  preliminary results]{OHC03}. Once velocity variations were
detected in these spectra, we searched both the \hst\ and
\fuse\ archives looking for any other observations. 
We found two sets of \fuse\ observations of Feige~48 in the archive,
the first (proposal no.\ B033, PI: G.\ Fontaine) was a companion
project to our \stis\ observations, and their goal was also to measure
metal abundances. The first results from this analysis is presented in
\citet{CFF03}. The second, larger set of data was part
of a project to study molecular gas in the interface between the
galactic disk and halo (proposal no.\ C036, PI: R.\ Benjamin). Results from
this investigation will be presented in a forthcoming paper
\citep{BWR04}. Fortunately for us, the star used to illuminate the gas cloud
under investigation was Feige~48. The list of observations are shown in Table
\ref{tab:obs}, and their description follows below.

\begin{table}
\caption{\hst\ and \fuse\ observation log for Feige 48.}
\label{tab:obs}
\begin{center}
\begin{tabular}{ccrc}
Date & UT & exp.\ time (s) & Instrument \\
\hline
2001-03-26 & 18:57:38 & 2662.828 & \fuse \\
2001-03-26 & 20:37:27 & 2567.906 & \fuse \\
2001-05-13 & 06:53:42 & 3600.200 & \stis \\
2001-05-13 & 07:59:32 & 3600.200 & \stis \\
2001-05-13 & 09:09:23 & 3500.199 & \stis \\
2001-05-13 & 10:13:33 & 3511.199 & \stis \\
2002-04-03 & 19:52:36 & 3856.992 & \fuse \\
2002-04-03 & 21:29:57 & 4226.070 & \fuse \\
2002-04-03 & 23:14:48 & 4181.148 & \fuse \\
2002-04-04 & 00:59:19 & 4264.227 & \fuse \\
2002-04-04 & 02:43:37 & 4379.312 & \fuse \\
\hline
\textbf{total:} & & 41175.457 &
\end{tabular}
\end{center}
\end{table}

\subsection{\stis\ UV echelle spectra}
\label{sec:stis}

We observed 5 sdBs with \stis: three pulsators (Feige~48,
PG\,1219+534 and PG\,1605+072) and two non pulsators (Feige 66 and
Ton\,S-227). The pulsators were all observed in time-tag mode, made
possible by the use of the MAMA detectors. This
was not originally intended, but was suggested (fortunately) by the time
allocation committee. Each star was observed using two medium
resolution echelle gratings, E140M and E230M, giving useful wavelength
ranges of 1150-1720\,\AA\ and 1630-2370\,\AA\ respectively. The slit
length and width used for all exposures was 0.2''$\times$0.06'', leading
to spectral resolutions of 0.017\,\AA\ at 1200\,\AA\ and 0.020\,\AA\ at
1500\,\AA\ with the E140M grating and 0.051\,\AA\ at 1700\,\AA\ and
0.068\,\AA\ at 2400\,\AA\ with the E230M grating. All spectra were
processed using the time-tag and standard reduction routines in the
\textsc{stsdas} package in \textsc{iraf}. The orders containing
Ly$\alpha$ were excluded, since they are not useful for a radial
velocity analysis.

\subsection{\fuse\ spectra}
\label{sec:fuse}

In both sets of observations, Feige~48 was observed using the
high-throughput 30''$\times$30'' LWRS aperture on all four
spectrograph channels and recorded in time-tag mode. All data were
reprocessed by the CalFUSE pipeline software (version 2.4). Of the
four channels available, only the LiF2 channel was used, since it
contains no Lyman lines which are affected by interstellar absorption
and airglow lines from the Earth. The detector also has the highest
sensitivity of all onboard \fuse. The dispersion of this channel is
6.7\,m\AA\,pixel$^{-1}$ and the wavelength range covered is
1086-1182\,\AA. For a description of the detectors and more
information on the \fuse\ mission, see \citet{FUSE2000}.

\section{Synthetic Spectra}
\label{sec:synthspec}

The UV spectra of Feige~48 (and sdBs in general) contain a large
number of spectral lines due to iron-group elements (V, Cr, Mn, Fe,
Co, Ni), making continuum definition often difficult. The number of lines
complicates abundance analyses, but is very
advantageous for measuring velocities, since they are very
sharp (rotation velocities in sdBs are typically $\le$5\,km\,s$^{-1}$).

As input for our spectrum synthesis, we used a metal line-blanketed LTE model
atmosphere with solar metallicity and Kurucz' ATLAS6 Opacity Distribution
Functions. The temperature, surface gravity, and helium abundance were set at
$T_{\mathrm{eff}}$=29\,500\,K, $\log g$=5.50, and log(He/H)=--2.93, as
determined by \citet{HRW00}.

The spectra were synthesised using Michael Lemke's version of the
LINFOR program (originally developed by Holweger, Steffen, and
Steenbock at Kiel University). Oscillator strengths were taken from
the Kurucz line list, as were damping constants for all metal
lines. Both the rotation and microturbulent velocities were set to
0\,km\,s$^{-1}$. (The rotation velocity of Feige~48 will be discussed
further in Section \ref{sec:nature}).

\section{Binary motion}
\label{sec:binary}

\begin{figure}
\vspace{6cm}
\begin{center}
    \includegraphics{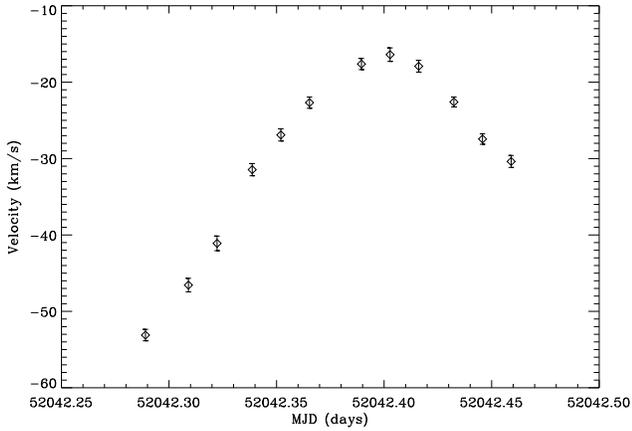}
  \end{center}
  \caption{Velocity variations from far and near UV spectra of Feige~48.}
  \label{fig:stisvel}
\end{figure}

During the analysis of abundances in Feige~48, we found that the
spectra from the two NUV orbits showed different velocity shifts
compared to our synthetic spectra. Because the spectra were taken in
time-tag mode, we were able to divide each of the far and near UV
exposures into three 1150\,s sub-exposures. These spectra were cross
correlated with our synthetic spectra and the results are shown in
Figure \ref{fig:stisvel}. The original \fuse\ observations were also
taken in time-tag mode, allowing a similar division, with the exposure
time typically being 1050\,s.

\begin{figure}
\vspace{6cm}
\begin{center}
    \includegraphics{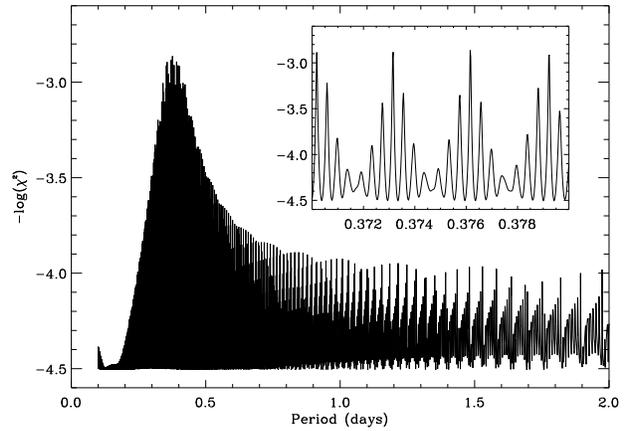}
  \end{center}
  \caption{Amplitude spectrum of the radial velocities of
  Feige~48. The inset shows the aliasing around the adopted period.}
  \label{fig:chi2spec}
\end{figure}

Using the double-precision cross-correlation routine (rvx) available
in \textsc{iraf},\footnote{available from
\texttt{http://iraf.noao.edu/extern/}} we measured the Doppler
shift of each newly-created spectrum relative to our synthetic
spectra. Since the \stis\ spectra are in echelle format, each
observation yielded 40 velocity measurements (one per order). We
calculated the weighted average of these velocities using the inverse
square of the velocity errors as weights. The end result of this was a
list of velocities with two large gaps in time. These velocities are
typically accurate to $\sim$1\,km\,s$^{-1}$.

The analysis of the radial velocity curve is made simpler because the
system is single-lined. By fitting the curve with a range of
periods we produced a ``power spectrum'', shown in Figure \ref{fig:chi2spec},
which shows the goodness of fit. The actual
period described below was found by selecting the highest peak in this
spectrum. There is
strong aliasing, leading to a relatively large period error (derived
by measuring the difference between the nearest alias peaks). 

\begin{figure*}
  \vspace{8cm}
  \begin{center}
    \includegraphics{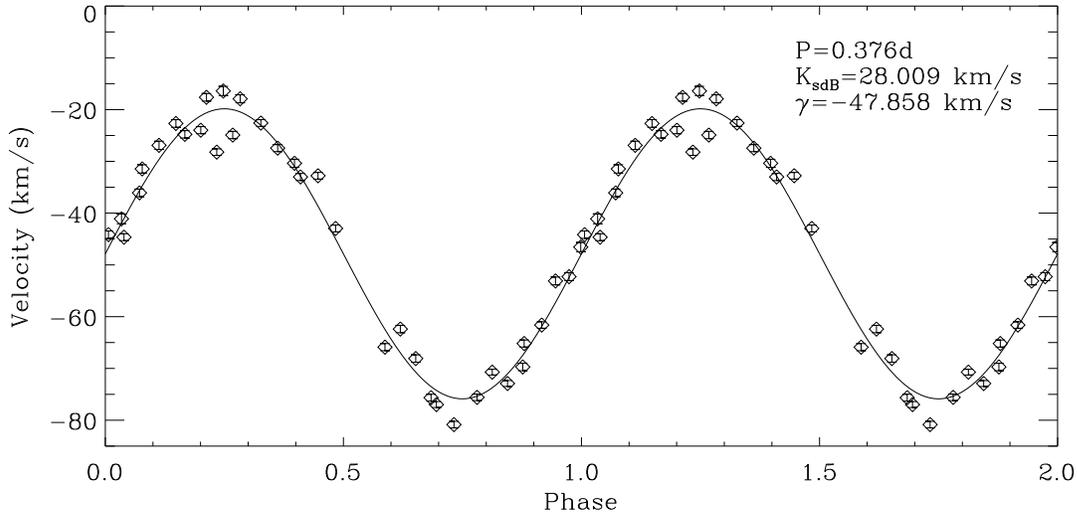}
  \end{center}
  \caption{The phased radial velocity curve of Feige~48. The errors
  shown are formal errors determined by \textsc{iraf}.}
  \label{fig:phasedvel}
\end{figure*}

The ephemeris of the system at time $T_0$, defined as the
conjunction time that the sdB moves from the blue to red side of the
velocity curve, is
\begin{eqnarray}
\mathrm{HJD}(T_0) & = & (2\,452\,367.05\pm0.03) \nonumber \\
 & & +(0.376\pm0.0003\pm0.003)\times E \nonumber
\end{eqnarray}
where the first period error quoted is derived from the uncertainty
due to the timespan of the observations, while the second error is derived
by measuring the difference between the nearest alias peaks, as discussed
above. We stress here that these values should not be taken as 1$\sigma$
uncertainties.
We measure the velocity semi-amplitude of Feige~48 to be
$K_{\mathrm{sdB}}$=28.0$\pm$0.2\,km\,s$^{-1}$, while the system
velocity is $\gamma$=--47.9$\pm$0.1\,km\,s$^{-1}$. The velocity
curve phased with this ephemeris is shown in Figure
\ref{fig:phasedvel}. Despite the separation in time of the
observations, the phase coverage is quite good. The period and
semi-amplitude of the sdB can be used to derive the mass function,
$f(m)$=0.000856$\pm$0.000019\,$M_\odot$. If we assume a mass of
$\sim$0.47\,$M_\odot$ for the sdB, as estimated from asteroseismology,
then the lower limit for the companion mass is 0.062\,$M_\odot$. Note
that the period we have derived here is well below the upper limit set
using the phase stability of the pulsations by \citet{RKZ04}. Their analysis
is insensitive to systems with short periods, small inclinations and massive
companions, however, since the light-travel time across the system is far too
small to cause a measureable effect on the pulsations.

\section{The nature of the companion}
\label{sec:nature}

In order to discuss the implications of this discovery for the
formation models of \citet{HanII}, we must first constrain the
nature of the companion to Feige~48.

\begin{figure}
  \vspace{6cm}
  \begin{center}
    \includegraphics{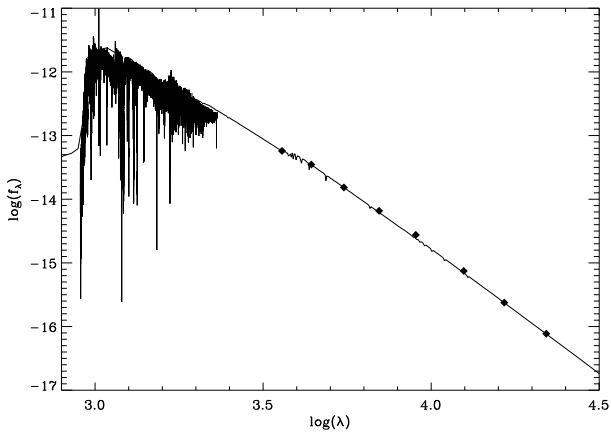}
  \end{center}
  \caption{Spectral energy distribution of Feige~48. The filled
    diamonds are Johnson magnitudes, including \emph{2MASS}
    magnitudes. The \stis\ and \fuse\ spectra are also
    shown. There is no evidence for the companion to the
    sdB.}
  \label{fig:f48sed}
\end{figure}

The small mass function implies three possible companion types for Feige~48: a
late-type main-sequence star, a brown dwarf or a white dwarf. We can place
limits on the first two types using multicolour photometry, as was done by
\citet{HELN03} for the sdB binary HD\,188112. Previous studies have suggested
that any main sequence companion must have spectral type later than mid K
\citep{ECpaperXI}. We can now compare the \emph{2MASS JHK} magnitudes, along
with other Johnson measurements available, with the model spectral energy
distribution of Feige~48, shown with photometric measurements overplotted in
Figure \ref{fig:f48sed}. Apart from \emph{2MASS}, the photometry is obtained
from a variety of sources \citetext{$U-B=-$1.03, $B-V=-$0.25, $V$ = 13.48 --
  \citealp{BW73}, $R$ = 13.57 (averaged), $I$ = 13.72 -- the USNO-B catalogue,
  \citealp{USNOB}}. Magnitudes were converted to fluxes using the calibration
of \citet{FSI95} for the optical and \citet{CWM03} for the infrared. The model
was then scaled to the $V$ flux of Feige~48, implying a distance to the system
of 755$\pm$90\,pc.  As can be seen in Figure \ref{fig:f48sed}, there is no
contribution to the flux from the companion in the $K$-band (on this scale,
the errors in $JHK$ magnitudes are smaller than the data points). Using the
absolute $K$ magnitude of dwarfs of spectral types M2-M9 determined by
\citet{BBH99}, we find that if the companion to Feige~48 is on the main
sequence, its spectral type must be M3 or later. This type of star will have a
mass of $\sim$0.35\,$M_\odot$, leading to a lower limit for the inclination
angle of $\sim$14$^\circ$.

\begin{figure*}
  \vspace{11cm}
  \begin{center}
    \includegraphics{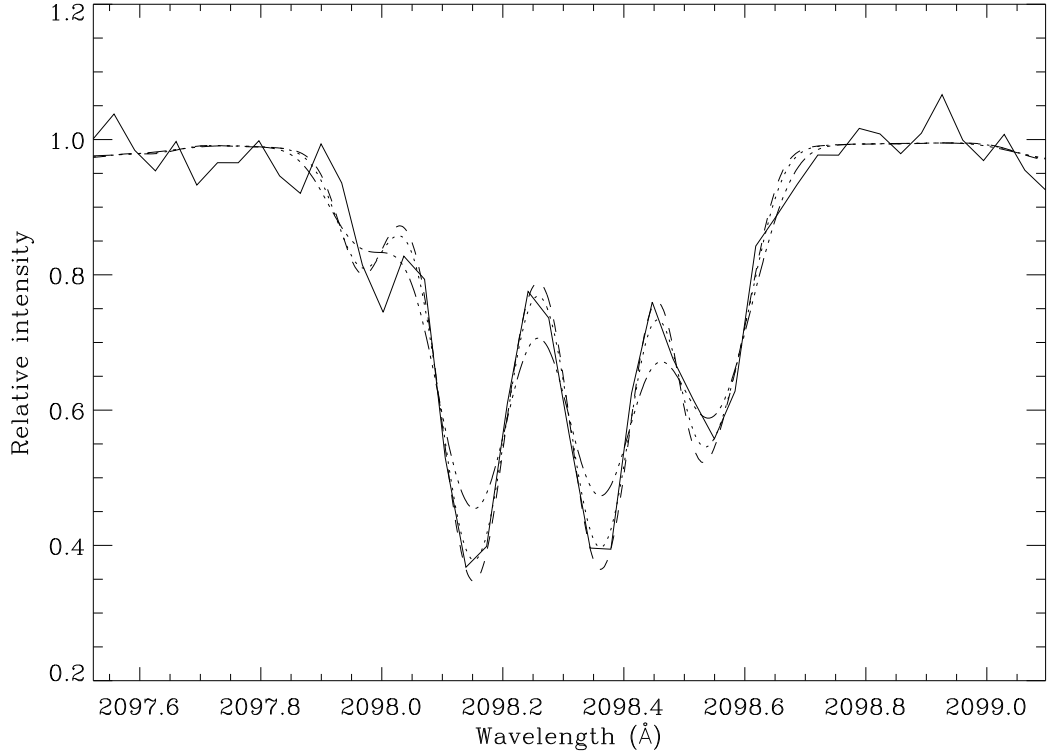}
  \end{center}
  \caption{Fit of $v\sin i$ to three Fe\,\textsc{III} lines. The line
    at $\sim$2098.5\,\AA\ is blended with a Mn\,\textsc{iii} line.
    The dashed line represents $v\sin i$=0\,km\,s$^{-1}$, the dotted
    line is $v\sin i$=5\,km\,s$^{-1}$, and the dash-triple-dot line is $v\sin
    i$=10\,km\,s$^{-1}$. See the text for further details.}
  \label{fig:vsinifit}
\end{figure*}

We can also consider information that has been obtained from the
pulsations themselves. \citet{MR_PhD02}, \citet{RKZ04} and \citet{CBF03} have
carried out asteroseismological analyses of Feige~48, and suggest a rotation
period of $\sim$9.5-10\,h. Given the uncertainties involved in the sdB
evolution models used and the aliasing seen in Figure \ref{fig:chi2spec}, this
is in very good agreement with the orbital period of 9.03\,h determined in
Section \ref{sec:binary}. In particular, both \citet{MR_PhD02} and
\citet{RKZ04} assumed a rotational splitting coefficient ($C_{nl}$) of
zero. If $C_{nl}$ is set to 0.1, which is not unreasonable for a slightly
evolved sdB, then the rotation period would be $\sim$9\,h.

We now compare the observed
upper limit on a reflection effect to what might be expected from an M3 dwarf
with radius $\sim$0.35\,$R_\odot$. A similar system, but with higher
inclination and shorter period is the sdB+dM HS\,2333+3927 \citep{HDOe04}. In
this case a large reflection effect is observed ($\Delta V$=0.28\,mag) even
though the inclination is only 36-39$^\circ$. This suggest that an effect
might be measureable for Feige~48 if present. Calculations by H.\ Drechsel
(private communication) find that an sdB+dM system with a period of 9\,h and
inclination 15$^\circ$ would show luminosity variations of 0.025-0.030\,mag
for maximal albedo of the secondary. The only time-series observations of
Feige~48 made were optimised for oscillation studies, not searching for a
reflection effect. With an orbital period of 9\,h, such an effect is difficult
to detect, since observations from a single site will typically not cover an
entire orbit. A re-analysis of these data places an upper limit of
$\sim$0.013\,mag on any reflection effect (M.\ Reed, private
communication). Given the uncertainties involved in these calculations and
comparisons, we do not feel confident completely ruling out an M3 dwarf
companion at this stage, but instead suggest the white dwarf scenario is more
likely.

Finally, using our high resolution UV spectra, we may be able to improve on
the equatorial rotation velocity limits set by \citet{HRW00} and therefore
constrain the orbital inclination and companion mass. In the near UV our
spectral resolution is almost a factor of two better than the spectra used by
\citeauthor{HRW00}, so we shifted each of the six spectra used to
determine radial velocities to their rest wavelength and then averaged
them. The spectrum was examined for metal lines in close blends
that are clearly resolved (a difficult task due to the very large number
of unresolved blends), and these lines were fit with a variety of values
of $v\sin i$. We show the three Fe\,\textsc{iii} lines chosen in Figure
\ref{fig:vsinifit}, with three different rotation velocities $v\sin i$=0,
5, and 10\,km\,s$^{-1}$. There is no clear difference between the
0\,km\,s$^{-1}$ (dashed) and 5\,km\,s$^{-1}$ curves (dotted), while $v\sin i$
= 10\,km\,s$^{-1}$ (triple-dash-dotted curve) does not appear to fit. After
a simple $\chi^2$ analysis, we find that $v\sin
i=4.8\pm1.2$\,km\,s$^{-1}$. At a 3$\sigma$ level $v\sin i$ is constrained to
less than 8\,km\,s$^{-1}$. Together with the upper limit of 5\,km\,s$^{-1}$
(3$\sigma$) derived from the Mg\,\textsc{ii} line in the optical
\citep{HRW00}, we consider 5\,km\,s$^{-1}$ as an upper bound the following
discussion.

The rotation of Feige~48 is expected to be tidally locked to the orbit. In
this case we can estimate the rotation velocity from the orbital period and
the radius ($\sim 0.2R_\odot$) to be
$26.9^{+1.5}_{-1.6}$\,km\,$^{-1}$ (this is coincidentally
  approximately the same as the orbital velocity). Comparison with the
projected rotational velocity constrained above yields an upper limit to the
inclination of the system of $i=11.4^\circ$. This is the same upper
limit derived by \citet{RKZ04} and \citet{MR_PhD02} when they combined the
upper limit to the rotation velocity derived by \citet{HRW00} and the
rotation period derived from their asteroseismological analysis. From the
mass function we derive a lower limit to the companion mass of
$M_2=0.46M_\odot$. This result means that the companion must be degenerate,
and is most likely a white dwarf. Considering the orbital period and companion
mass, the Feige~48 system should have gone through two common envelope phases,
and therefore can be explained by the ``second common envelope ejection
channel'' described by \citet{HanII}.

\section{Discussion and Conclusions}
\label{sec:disc}

We have detected binary motion in the V361\,Hya star Feige~48 with a
period of 0.376\,d and a velocity semi-amplitude of
28.0\,km\,s$^{-1}$. Combining multicolour photometry, the
asteroseismological determination of the rotation period, and an
upper limit of the equatorial rotation velocity, we conclude
that the orbital inclination $i\le$11.4$^\circ$. This implies that the minimum
mass of the companion is 0.46\,$M_\odot$; combining this with the lack of a
reflection effect suggests the star must be a white dwarf.

Feige 48 is one of only two pulsating sdBs with a known white dwarf
companion. Unlike KPD\,1930+2752, the other system, it has a relatively simple
pulsation spectrum and no (or very little) ellipsoidal distortion. This has
allowed \citet{MR_PhD02}, \citet{RKZ04} and \citet{CBF03} to match pulsation
frequencies and constrain sdB evolution models. However, these studies assumed
that Feige 48 evolved according to canonical EHB theory. While this is
probably accurate since the star has undergone two phases of common envelope
evolution, the conclusions of \citet{HanII} suggest that investigating a wider
mass range than previously may be warranted, with masses as low as
$\sim$0.36\,$M_\odot$. In the case of single sdBs, which should be formed
through the merger channel, masses in the range 0.4-0.7\,$M_\odot$ should be
examined. The sdBs in long-period orbits with main sequence companions (formed
through the Roche lobe overflow channel) may fall in the 0.3-0.5\,$M_\odot$
range, and possibly even higher, although \citeauthor{HanII} find this to be
unlikely.  Enlarging the mass range investigated in this way may provide
another useful test of the proposed formation models for sdBs.

These tests can only be carried out once the binary nature of the pulsating
sdBs is known. Up to now only
seven V361\,Hya stars are confirmed binaries; four have F- or G-type main
sequence companions (V361\,Hya itself, PB\,8783, EC\,20117-4014 and
EC\,10228-0905), two have white dwarf companions (KPD\,1930+2752 and
Feige~48), and one is in an eclipsing binary system with an M5 main
sequence star (PG\,1336--018). That leaves around 25 pulsators with
unknown status. The fraction of sdBs is binary systems is still under
debate, however if we consider the best-fit simulation of \citet{HanII}, we
find that $\sim$41\% or 10-11 of these stars should have either
late-type main sequence or white dwarf companions.
\citet{Lisker2004} compared their observations of 52 sdB stars with the 12
simulation sets constructed by \citet{HanII} using a simple statistical test.
The simulation set they found to fit their observations differs from that
found qualitatively by \citeauthor{HanII} and suggests that up to $\sim$53\%
or $\sim$13 of the pulsating sdBs should have an unseen companion.
\emph{High precision spectra are needed to observe
  short period systems such as Feige~48, so a survey with high
  spectral and temporal resolution is required.}

\begin{acknowledgements}

We would like to thank Mike Reed and Horst Drechsel for
useful discussions.  SJOT is supported by the Deutsches Zentrum f\"ur Luft- und
Raumfahrt (DLR) through grant no.\ 50-OR-0202. RAB acknowledges support from
the \emph{FUSE} guest observer program (NAG 5-12196). This publication makes
use of data products from the Two Micron All Sky Survey, which is a joint
project of the University of Massachusetts and the Infrared Processing and
Analysis Center/California Institute of Technology, funded by the National
Aeronautics and Space Administration and the National Science Foundation.
\end{acknowledgements}

\bibliographystyle{aa}

\end{document}